# Research Agenda in Cloud Technologies

Ilango Sriram

Department of Computer Science
University of Bristol
Bristol, UK

ilango@cs.bris.ac.uk

Ali Khajeh-Hosseini

Cloud Computing Co-laboratory
School of Computer Science
University of St Andrews
St Andrews, UK

akh@cs.st-andrews.ac.uk

## ABSTRACT
Cloud computing is the latest effort in delivering computing resources as a service. It represents a shift away from computing as a product that is purchased, to computing as a service that is delivered to consumers over the internet from large-scale data centres – or "clouds". Whilst cloud computing is gaining growing popularity in the IT industry, academia appeared to be lagging behind the rapid developments in this field. This paper is the first systematic review of peer-reviewed academic research published in this field, and aims to provide an overview of the swiftly developing advances in the technical foundations of cloud computing and their research efforts. Structured along the technical aspects on the cloud agenda, we discuss lessons from related technologies; advances in the introduction of protocols, interfaces, and standards; techniques for modelling and building clouds; and new use-cases arising through cloud computing.

## Categories and Subject Descriptors
A.1 [**General Literature**]: Introductory and Survey
C.2.4 [**Computer Communication Networks**]: Distributed Systems – *Cloud Computing*

## 1. INTRODUCTION
Cloud computing has recently reached popularity and developed into a major trend in IT. While industry has been pushing the Cloud research agenda at high pace, academia has only recently joined, as can be seen through the sharp rise in workshops and conferences focussing on Cloud Computing. Lately, these have brought out many peer-reviewed papers on aspects of cloud computing, and made a systematic review necessary, which analyses the research done and explains the resulting research agenda. We performed such a systematic review of all peer-reviewed academic research on cloud computing, and explain the technical challenges facing in this paper.

There were several whitepapers and general introductions to cloud computing, which provide an overview of the field, [e.g. 1, 2, 3, 4, 5], but yet there is no systematic review of the agenda academia has taken. Pastaki Rad *et al.* [6] presented a preliminary survey that included a short overview of storage systems and Infrastructure as a Service (IaaS), which, however, was not systematic and fell short of providing a good overview of the state-of-the-art and lacked a discussion of the research challenges. Our paper aims to provide a comprehensive review of the academic research done in cloud computing and to highlight the research agenda academia is pursuing. We are well aware that a survey in such a fast moving field will soon be out of date, but feel such a survey would provide a good base for the 1st ACM Symposium on Cloud Computing to set new work in context with, and that it can act as a resource for researchers new in this area. Research in this field appeared to be split into two distinct viewpoints. One investigates the technical issues that arise when building and providing clouds, and the other looks at implications of cloud computing on enterprises and users. In this paper we discuss the advances and research questions in technical aspects of Cloud Computing, such as protocols, interoperability and techniques for building clouds, while we discuss the research challenges facing enterprise users, such as cost evaluations, legal issues, trust, privacy, security, and the effects of cloud computing on the work of IT departments, elsewhere [7]. This paper is structured as follows: the methodology used to carry out this review is shown in the Section 2; Section 3 discusses various definitions of cloud computing; Section 4 outlines the lessons to be learnt from related areas; Section 5 and Section 6 review the work on standardised interfaces and Cloud interoperability respectively; Section 7 summarises various other research done in support of building Cloud infrastructures; while use cases of Cloud computing are reviewed in Section 8; finally Section 9 concludes the review by summing up the research directions academia faces.

## 2. METHODOLOGY
This review surveyed the existing literature using a principled and systematic approach: we searched each of the major research databases for computer science, the ACM Digital Library, IEEE Xplore, SpringerLink, ScienceDirect and Google Scholar, for the following keywords: cloud computing, elastic computing, utility computing, Infrastructure as a Service, IaaS, Platform as a Service, PaaS, Software as a Service, SaaS, Everything as a Service, XaaS. The date range for this search was limited from 2005 until October 2009. This date range was chosen because this survey work was commenced in October 2009, and because all public clouds were launched after 2005. For example, Amazon first launched EC2 (Elastic Compute Cloud) in August 2006[1] and Google launched App Engine in April 2008[2]. According to Google Trends, the term *cloud*

---
[1] http://aws.typepad.com/aws/2006/08/amazon_ec2_beta.html

[2] http://googleappengine.blogspot.com/2008/04/introducing-google-app-engine-our-new.html

*computing* started becoming popular in 2007 as shown in Figure 1.

The searches from the five target databases returned over 150 papers. The titles and abstracts of these papers were read and for quality reasons we decided to use only peer-reviewed papers for the review; only a small number of non peer-reviewed publications were included, such as well quoted definitions or a summary of a workshop discussing research challenges academia is facing, as these were relevant and not matched by comparable peer-reviewed work. Furthermore, papers that had misleading titles or abstracts and those that were purely focused on High Performance Computing and e-Science were also left out of the review as these areas are not within the core focus of our review. The citation-references of the selected papers were checked but no additional papers were found to be necessary to add to this review based on the criteria mentioned above. This resulted in a total of 56 publications being selected for review. The papers were split into three categories based on their main focus; the categories were: general introductions, technological aspects of cloud computing and organizational aspects. The latter category is discussed elsewhere [7]. The papers that provided general introductions to cloud computing are referenced throughout this paper. The technological category was further broken down into papers that dealt with protocols, interfaces, standards, lessons from related technologies, techniques for modelling and building clouds, and new use-cases arising through cloud computing.. Table 1 provides an overview of the papers reviewed in this review and their categories. As it can be seen in the table, the majority of the papers were published in 2009.

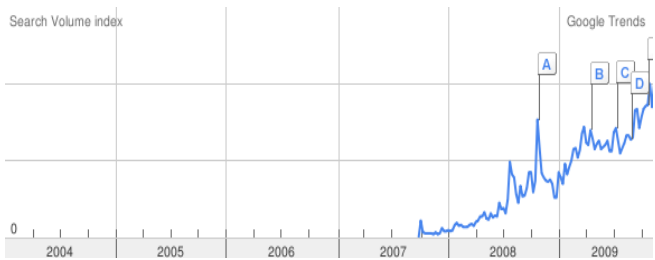

**Figure 1: Searches for "cloud computing" on Google.com, taken from Google Trends**

## 3. DEFINITIONS

There has been much discussion in industry as to what cloud computing actually means. The term cloud computing seems to originate from computer network diagrams that represent the internet as a cloud. Most of the major IT companies and market research firms such as IBM [8], Sun Microsystems [1], Gartner [9] and Forrester Research [10] have produced whitepapers that attempt to define the meaning of this term. These discussions are mostly coming to an end and a common definition is starting to emerge. The US National Institute of Standards and Technology (NIST) has developed a working definition that covers the commonly agreed aspects of cloud computing. The NIST working definition summarises cloud computing as:

*a model for enabling convenient, on-demand network access to a shared pool of configurable computing resources (e.g., networks, servers, storage, applications, and services) that can be rapidly provisioned and released with minimal management effort or service provider interaction* [11]

**Table 1: Overview of the reviewed literature**

| Category | Authors |
|---|---|
| General introductions | Armbrust et al. 2009, Carr 2008, Erdogmus 2009, Foster et al. 2008, Pastaki Rad et al. 2009, Voas and Zhang 2009, Vouk 2008 |
| Definitions | Mell and Grance 2009, Vaquero et al. 2009, Youseff et al. 2008 |
| Protocols, interfaces, and standards | Bernstein et al. 2009, Dodda et al. 2009, Grossman 2009, Harmer et al. 2009, Keahey 2009, Lim et al. 2009, Matthews et al. 2009, Mikkilineni and Sarathy 2009, Nurmi et al. 2008, Ohlman et al. 2009, Sun et al. 2007 |
| Lessons from related technologies | Buyya et al. 2008, Chang 2006, Foster et al. 2008, Napper and Bientinesi 2009, Sedayao 2008, Vouk 2008, Zhang and Zhou 2009 |
| Building clouds | AbdelSalam et al. 2009, Buyya et al. 2009, Song et al. 2009, Sotomayor et al. 2009, Sriram 2009, Vishwanath et al. 2009 |
| Use cases | Chun and Maniatis 2009, Ganon and Zilbershtein 2009, Matthew and Spraetz 2009, Wilson 2009 |

The NIST definition is one of the clearest and most comprehensive definitions of cloud computing and is widely referenced in US government documents and projects. This definition describes cloud computing as having five essential characteristics, three service models, and four deployment models. The essential characteristics are:

- On-demand self-service: computing resources can be acquired and used at anytime without the need for human interaction with cloud service providers. Computing resources include processing power, storage, virtual machines etc.

- Broad network access: the previously mentioned resources can be accessed over a network using heterogeneous devices such as laptops or mobiles phones.

- Resource pooling: cloud service providers pool their resources that are then shared by multiple users. This is referred to as *multi-tenancy* where for example a physical server may host several virtual machines belonging to different users.

- Rapid elasticity: a user can quickly acquire more resources from the cloud by scaling out. They can scale back in by releasing those resources once they are no longer required.

- Measured service: resource usage is metered using appropriate metrics such monitoring storage usage, CPU hours, bandwidth usage etc.

The above characteristics apply to all clouds but each cloud provides users with services at a different level of abstraction, which is referred to as a service model in the NIST definition. The three most common service models are:

- Software as a Service (SaaS): this is where users simply make use of a web-browser to access software that others have developed and offer as a service over the web. At the SaaS level, users do not have control or access to the underlying infrastructure being used to host the software. Salesforce's Customer Relationship Management software[3] and Google Docs[4] are popular examples that use the SaaS model of cloud computing.

- Platform as a Service (PaaS): this is where applications are developed using a set of programming languages and tools that are supported by the PaaS provider. PaaS provides users with a high level of abstraction that allows them to focus on developing their applications and not worry about the underlying infrastructure. Just like the SaaS model, users do not have control or access to the underlying infrastructure being used to host their applications at the PaaS level. Google App Engine[5] and Microsoft Azure[6] are popular PaaS examples.

- Infrastructure as a Service (IaaS): this is where users acquire computing resources such as processing power, memory and storage from an IaaS provider and use the resources to deploy and run their applications. In contrast to the PaaS model, the IaaS model is a low level of abstraction that allows users to access the underlying infrastructure through the use of virtual machines. IaaS gives users more flexibility than PaaS as it allows the user to deploy any software stack on top of the operating system. However, flexibility comes with a cost and users are responsible for updating and patching the operating system at the IaaS level. Amazon Web Services' EC2 and S3[7] are popular IaaS examples.

Erdogmus [12] described Software as a Service as the core concept behind cloud computing, suggesting that it does not matter whether the software being delivered is infrastructure, platform or application, "it's all software in the end" [12]. Although this is true to some extent, it nevertheless helps to distinguish between the types of service being delivered as they have different abstraction levels. The service models described in the NIST definition are deployed in clouds, but there are different types of clouds depending on who owns and uses them. This is referred to as a cloud deployment model in the NIST definition and the four common models are:

---

[3] http://www.salesforce.com/uk/crm/products.jsp

[4] http://docs.google.com

[5] http://code.google.com/appengine

[6] http://www.microsoft.com/windowsazure/

[7] http://aws.amazon.com/

- Private cloud: a cloud that is used exclusively by one organisation. The cloud may be operated by the organisation itself or a third party. The St Andrews Cloud Computing Co-laboratory[8] and Concur Technologies [13] are example organisations that have private clouds.

- Public cloud: a cloud that can be used (for a fee) by the general public. Public clouds require significant investment and are usually owned by large corporations such as Microsoft, Google or Amazon.

- Community cloud: a cloud that is shared by several organisations and is usually setup for their specific requirements. The Open Cirrus cloud testbed could be regarded as a community cloud that aims to support research in cloud computing [14].

- Hybrid cloud: a cloud that is setup using a mixture of the above three deployment models. Each cloud in a hybrid cloud could be independently managed but applications and data would be allowed to move across the hybrid cloud. Hybrid clouds allow cloud bursting to take place, which is where a private cloud can burst-out to a public cloud when it requires more resources.

Figure 2 provides an overview of the common deployment and service models in cloud computing, where the three service models could be deployed on top of any of the four deployment models.

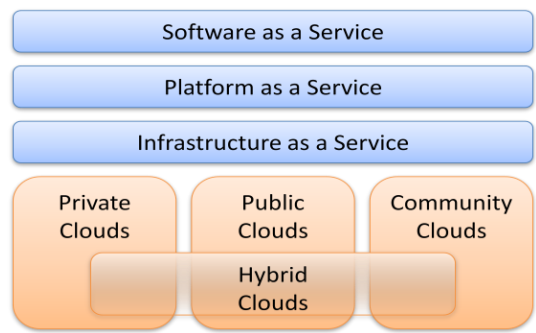

**Figure 2: Cloud computing deployment and service models**

Others such as Vaquero *et al.* [15] and Youseff *et al.* [16] concur with the NIST definition to a significant extent. For example, Vaquero *et al.* studied 22 definitions of cloud computing and proposed the following definition:

*Clouds are a large pool of easily usable and accessible virtualized resources (such as hardware, development platforms and/or services). These resources can be dynamically re-configured to adjust to a variable load (scale), allowing also for optimum resource utilization. This pool of resources is typically exploited by a pay-per-use model in which guarantees are offered by the Infrastructure Provider by means of customized SLAs.*

This definition includes three of the five characteristics of cloud computing described by NIST, namely resource pooling, rapid elasticity and measured service but fails to mention on-demand self-service and broad network access. Youseff *et al.* [16] described a five-layer stack that can be used to classify cloud services; they use composability as their methodology where each service is composed of other services. The five layers are applications, software environment, software infrastructure, software kernel, and hardware. This is similar to the SaaS, PaaS and IaaS service models described in the NIST definition and only differs in the lower two layers, namely the software kernel and hardware layers. Grid and cluster computing systems such as Globus and Condor are examples of cloud services that fall into the software kernel layer, and ultra large-scale data centres as designed in IBM's Kittyhawk Project [17] are examples of hardware layer services [16]. However, these are not convincing examples of cloud services as they do not have the essential characteristics of cloud computing as described in the NIST definition, therefore we feel that the two extra layers used by Youseff *et al.* could reasonably be seen as unnecessary when describing cloud computing.

---

[8] http://www.cs.st-andrews.ac.uk/stacc

It is useful to think of a *cloud* as a collection of hardware and software that runs in a data centre and enables the cloud computing model [18]. "Scalability, reliability, security, ease of deployment, and ease of management for customers, traded off against worries of trust, privacy, availability, performance, ownership, and supplier persistence" are the benefits of cloud computing for Erdogmus [12].

Although there are still many internet forum and blog discussions on what cloud computing is and is not, the NIST definition seems to have captured the commonly agreed aspects of cloud computing that are mentioned in most of the academic papers published in this area. However, cloud computing is still in its infancy and as acknowledged by the authors Mell and Grance [11], this and any definition is likely to evolve in the future as new developments in cloud computing are explored. The current two-page NIST definition of cloud computing could be nicely summarised using Joe Weinman's retro-fitted *CLOUD* acronym that describes a cloud as a **C**ommon, **L**ocation-independent, **O**nline **U**tility provisioned on-**D**emand [19].

## 4. Lessons from related Technologies

The remainder of this paper reviews the research that describes technological aspects of research in cloud computing. This starts with a look at lessons to be learnt from related fields of research. In the following, standards and interfaces in cloud computing as well as interoperability between different cloud systems are explained. Then, techniques for designing and building clouds are summarised, which include advances in management software, hardware provisioning, and simulators that have been developed to evaluate design decisions and cloud management choices. This is rounded up by presenting new use-cases that have become possible through cloud computing.

Voas and Zhang [20] identified cloud computing as the next computing paradigm that follows on from mainframes, PCs, networked computing, the internet and grid computing. These developments are likely to have similarly profound effects as the move from mainframes to PCs had on the ways in which software was developed and deployed. One of the reasons that prevented grid computing from being widely used was the lack of virtualization that resulted in jobs being dependant on the underlying infrastructure. This often resulted in unnecessary complexity that had an effect on wider adoption [21]. Ian Foster – who was one of the pioneers of grid computing – compared cloud computing with grid computing and concluded that although the details and technologies of the two are different, their vision is essentially the same [22]. This vision is to provide computing as a utility in the same way that other public utilities such as gas and electricity are provided. In fact the dream of utility computing has been around since the 1960s and advocated by the likes of John McCarthy and Douglas Parkhill. For example, the influential mainframe operating system *Multics* had a number of design goals that are remarkably similar to the aims of current cloud computing providers. These design goals included remote terminal access, continuous operational provision (inspired by electricity and telephone services), scalability, reliable file systems that users trust to store their only copy of files, information sharing controls, and an ability to support different programming environments [23]. Therefore it is unsurprising that many people compare cloud computing to mainframe computing. However, it should be noted that although many of the ideas are the same, the user experience of cloud computing is almost completely the opposite of mainframe computing. Mainframe computing limited people's freedom by restricting them to a very rigid environment; cloud computing expands their freedom by giving them access to a variety of resources and services in a self-service manner.

Foster *et al.* [22] compare and contrast cloud computing with grid computing. They believe cloud computing is an evolved version of grid computing, in such a way that it answers the new requirements of today's time, takes into account the expensiveness of running clusters, and the existence of low-cost virtualisation. IT has greatly evolved in the last 15 years since grid computing was invented, and at present it is on a much larger scale that enables fundamentally different approaches. Foster *et al.* see similarities between the two concepts in their vision and architecture, see a relation between the concepts in some fields as in the programming model ("MapReduce is only yet another parallel programming model") and application model (but clouds are not appropriate for HPC applications that require special interconnects for efficient multi-core scaling), and they explain fundamental differences in the business model, security, resource management, and abstractions. Foster *et al.* find that in many of these fields there is scope for both the cloud and grid research communities to learn from each other's findings, and highlight the need for open protocols in the cloud, something grid computing adopted in its early days. Finally, Foster *et al.* believe that neither the electric nor computing grid of the future will look like the traditional electric grid. Instead, for both grids they see a mix of micro-productions (alternative energy or grid computing) and large utilities (large power plants or data centres).

In Market-Oriented Cloud Computing, a follow-on work from their Market-Oriented Grid Computing and Market-Oriented Utility Computing papers, Buyya *et al.* [24] describe their work on market oriented resource allocation and their Aneka resource broker: In the case of limited availability of resources, not all service requests will be of equal importance, and a resource broker will regulate the supply and demand of resources at market equilibrium. A batch job for example might be preferably processed when the resource value is low, while a critical live service request would need to be processed at any price. Aneka, commercialised through Manjrasoft, is a service broker that mediates between consumers and providers by buying capacities from the provider and subleasing them to the consumers. However, such resource trading requires the availability of ubiquitous cloud platforms with limited resources, and is in contrast to the desire for simple pricing models.

As cloud computing delivers IT as a service, cloud researchers can also learn from service oriented architecture (SOA). In fact, the first paper that introduced PaaS [25] described PaaS as an artefact of combining infrastructure provisioning with the principles of SaaS and SOA. Since then, no academic work has been published in the field of PaaS. We have to take our to-date understanding of PaaS from the current developments in industry, in particular from the two major vendors, Force.com and from Google App Engine. Sedayao [26] built a monitoring tool using SOA services and principles, and describe their experience from building a robust distributed application consisting of unreliable parts and the implication for cloud computing. As design goal for distributed computing scenarios such as cloud computing they propose, "like routers in a network, any service using other cloud services needs to validate input and have hold down periods before determining that a service is down"[26]. Zhang and Zhou [27] analyse convergence from SOA and virtualisation for cloud computing and present seven architectural principles and derive ten interconnected architectural modules. These build the foundation for their IBM cloud usage model, which is proposed as Cloud Computing Open Architecture (CCOA). Vouk [21] described cloud computing from a SOA perspective and talked about the Virtual Computing Laboratory (VCL) as an implementation of a cloud. VCL is an "open source implementation of a secure production-level on-demand utility and service oriented technology for wide-area access to solutions based on virtualised resources, including computational, storage and software resources" [21]. In this respect, VCL could be categorised as an IaaS layer service.

Napper and Bientinesi [28] ran an experiment to compare the potential performance of Amazon's cloud computing with the performance of the most powerful, purpose build, high performance computers (HPC) in the Top500 list in terms of solving scientific calculations using the LINPACK benchmark. They found that the performance of individual nodes in the cloud is similar to those in HPC, but that there is a severe loss in performance when using multiple nodes, although the used benchmark was expected to scale linearly. The AMD instances scaled significantly better than the Intel instances, but the cost for the computations were equivalent with both types. As the performance achieved decreased exponentially in the cloud and only linearly in HPC systems, Napper and Bientinesi [28] conclude that despite the vast availability of resources in cloud computing, these offerings are not able to compete with the supercomputers in the Top500 list for scientific computations.

In a non peer-reviewed summary of keynote speeches for a workshop on distributed systems Birman *et al.* [29] express that the distributed systems research agenda is quite different to the cloud agenda. They argue that while technologies from distributed systems are relevant for cloud computing, they are no longer central aspects of research. As example they list strong synchronisation and consistency as ongoing research topics from distributed systems. In cloud computing they remain relevant, but as the overarching design goal in the cloud is scalability, the search is now for decoupling and thus avoiding synchronisation, rather than improving synchronisation technologies. Birman *et al.* [29] come to a cloud research agenda comprising four directions: managing the existing compute power and the loads present in the data centre; developing stabile large-scale event notification platforms and management technologies; improving virtualisation technology; and understanding how to work efficiently with a large number of low-end and faulty components.

Cloud computing has been compared to several related fields of research. This section has shown that the cloud computing research agenda differs from the agenda in related fields, but that there are several findings in related research communities the research community can benefit from. We have also seen, that practitioners in distributed computing, grid computing, and SOA have joined the cloud community and proposed goals for research based on the background of their field. In the following, we shall look at the research more from the point of view of the cloud agenda.

## 5. STANDARDS AND INTERFACES

Cloud computing seeks to be a utility delivered in a similar as way electricity is delivered. Due to the higher complexity involved in delivering IT resources, open standards are necessary that enable an open market of providing and consuming resources. Currently, each vendor develops its own solution and avoids too much openness, to tie consumers in to their services and make it hard for them to switch to competitors. However, to new adopters the fear of vendor lock-in presents a barrier to cloud adoption and increases the required trust. There are three groups currently working on standards for cloud computing: The Cloud Computing Interoperability Forum[9], the Open Cloud Consortium[10], and the DMTF Open Cloud Standards Incubator[11]. There is also a document called the open cloud manifesto[12], in which various stakeholders express why open standards will benefit cloud computing. In literature, Grossman [2009] points out that the current state of standards and interoperability in cloud computing is similar to the early Internet era where each organization had its own network and data transfer was difficult. This changed with the introduction of TCP and other Internet standards. However, these standards were initially resisted by vendors just as standardisation attempts in cloud computing are being resisted by some vendors.

Keahey *et al.* [30] looked into the difficulties of developing standards and summarised the main goals of achieving interoperability between different IaaS providers as being *machine-image* compatibility, *contextualization* compatibility and *API-level* compatibility. Image compatibility is an issue as there are multiple incompatible virtualisation implementations such as the Xen, KVM, and VMWare hypervisors. When users want to move entire VMs between different IaaS providers, from the technological point of view this can only work when both providers use the same form of virtualisation. Contextualization compatibility problems exist because different IaaS providers use different methods of customizing the context of VMs, for example setting the operating system's username and password for access after deployment must be done in different ways. Finally, there are no widely agreed APIs between different IaaS providers that can be used to manage virtual infrastructures and access VMs. For machine image or VM compatibility there is an ongoing attempt to create an open standard called the Open Virtual Machine Format (OVF). At the API-level, for PaaS AppScale[13], an open source effort to re-implement the interfaces of Google App Engine, is aiming to become a standard, and for IaaS management, Amazon EC2's APIs are quickly becoming a *de-facto* standard, popularised through their open source re-implementation Eucalyptus.

---

[9] http://www.cloudforum.org

[10] http://www.opencloudconsortium.org

[11] http://www.dmtf.org/about/cloud-incubator

[12] http://www.opencloudmanifesto.org

Eucalyptus is an open-source software package that can be used to build IaaS clouds from computer clusters [31]. Eucalyptus emulates the proprietary Amazon EC2 SOAP and Query interface, and thus an IaaS infrastructure set up using Eucalyptus can be controlled with the same tools and software that is used for EC2. The open source nature of Eucalyptus gives the community a useful research tool to experiment with IaaS provisioning. The initial version of Eucalyptus used Xen as hypervisor for virtual machines, but since the publication of that version, support for further hypervisors has been added, in particular for the newly popular KVM hypervisor[14]. Eucalyptus has a hierarchical design that makes it reasonably easy to predict its performance. However, for very large data centres this centralised design might not scale particularly well, hence Nurmi *et al.* recommend it for typical settings in present in academia. Although Eucalyptus just re-implemented the Amazon EC2 interfaces, to date it is one of the most fundamental contributions by the research community towards standards in cloud computing, although only a few other providers use these interface APIs yet. But, for reasons such as fault tolerance or performance, or freedom from lock-in, consumers may wish to use multiple cloud providers. In the absence of open standards, or when attempts at providing open interface standards like Eucalyptus are not followed by some providers, there will be heterogeneous interfaces. Dodda *et al.* [32] address the problem of managing cloud resources with such heterogeneous access, by proposing a generic interface to the specific interface presented by individual cloud providers. They use their interface to an interface to compare the performance of Amazon EC2's Query and SOAP interface, and find that the average response time for the SOAP interface was nearly double that of the Query interface. These results emphasise the importance of selecting the interface through which resources from a given provider are managed. In a similar effort, Harmer *et al.* [33] present a cloud resource interface that hides the details of individual APIs to allow provider agnostic resource usage. They present the interface to create a new instance at Amazon EC2, at Flexiscale[15], and at a provider of on-demand non-virtualised servers called NewServers[16], and implemented an abstraction layer for these APIs. The solution from Harmer *et al.* goes beyond hiding API details and contains functionality to compensate for loss of core infrastructure in scenarios where multiple providers are used.

Cloud computing can benefit from standardised API interfaces as generic tools that manage cloud infrastructures can be developed for all offerings. For IaaS there are developments towards standards and Eucalyptus is looking to become the *de-facto* standard. For PaaS and SaaS stakeholders need to join the standardisation groups to work towards it. Achieving standardised APIs appears to be rather politically than technically challenging, hence there seems to be little space for academic involvement. However, standardised interfaces alone do not suffice to prevent vendor lock-in. For an open cloud, there is a need for protocols and software artefacts that allow interoperability to unlock more of the potential benefits from cloud computing. This technically rich direction will be discussed in the following section.

## 6. CLOUD INTEROPERABILITY AND NOVEL PROTOCOLS

The next steps from compatible and standardised interfaces towards utility provisioning are universal open and standard protocols that allow interoperability between clouds and enable the use of different offerings for different use cases. Bernstein *et al.* [34] describe an in-depth overview of the technological research agenda and open questions for interoperability in the cloud. They are looking for ways of allowing cloud services to interoperate with other clouds and highlight many goals and challenges, such as that cloud services should be able to implicitly use others through some form of library without the need to explicitly reference them, e.g. with their domain name and port. The collection of protocols inside and in-between the clouds that solve interoperability in the cloud are termed *intercloud protocols*. The intercloud protocol research agenda is made up of several areas: addressing, naming identity and trust, presence and messaging, virtual machines, multicast, time synchronisation, and reliable application transport. For cloud computing, each of these areas contains several issues. In addressing for example, the research problem is that there is the limited address space in IPv4 and that its successor IPv6 might be an inappropriate approach in a large and highly virtualised environment, as the cloud, due to its static addressing scheme: Bernstein *et al* criticise that IP addresses traditionally embody network locations for routing purposes and identity information, but in the cloud context identifiers should allow the objects to move into different subnets dynamically. This problem of static addresses is addressed by Ohlman *et al.* [35]. They recommend the usage of Networking of Information (NetInf) for cloud computing systems. Unlike URLs which are location-dependent, NetInf uses a location-independent model of naming objects, and offers an API that hides the dynamics of object locations and network topologies. Ohlman *et al.* demonstrate how this can ease management in the cloud, where the design desires transparency of location.

Further desired interoperability developments are listed by Matthews *et al.* [36], who propose virtual machine contracts (VMC) as an attempt at standardising VM protection and security settings, and are working on adding VMCs to the Open Virtual Machine Format (OVF) as extension to the metadata. Even in data centres under automated control and management, it is necessary to have customised settings for security and protection, such as firewall rules and bandwidth allowances for individual VMs. Today, these settings usually require the virtual appliance designer to manually communicate to both the person deploying the VM and to the administrators of the system, and the settings are specified and communicated in a site-specific non-portable format. In addition to allowing automated data centre management and cloud interoperability, Matthews *et al.* list several use cases for VMCs: Support for examining migration of enterprise data centres to the cloud; setting bounds on resource consumption and allowing capacity planning; detecting compromised VMs by comparing the VMs behaviour to the specified resource consumption estimates; specifying virtual network access control; specifying rules that ensure regulatory compliance and ease auditing compliance; and supporting disaster recovery as the required infrastructure elements will be known without having to instantiate a copy of the VM on a recovery cluster. Another piece of work that looks into cloud interoperability is Lim *et al.*'s [37] feedback control service for scaling in the cloud. Lim *et al.* say scaling choices must be under control of the users, in order to have control over spending and to be able to work towards maximising return on investments. Thus, feedback control systems that make scaling decisions need to be decoupled from the cloud provider. In experiments using CPU utilisation as threshold for scaling choices, the best results were found when coarse grained ranges where specified as desired states. Lim *et al.* intend to consider further sensors such as application level metrics of queue lengths or response times for scaling choices in their future work, and as open research question they ask how much internal cloud knowledge such a controller would minimally need in order to be effective, and how much control needs to be exposed by the cloud for this.

---

[13] http://code.google.com/p/appscale

[14] http://www.linux-kvm.org

[15] http://www.flexiscale.com

[16] http://www.newservers.com

Sun *et al.* [38] have looked into the integration of SaaS services. They argue that complete or full-blown solutions are too costly and hard to configure, so that real applications will require an integration of multiple individual SaaS products. Further, they see a need for SaaS products to be seemlessly integrated with user's existing in-house applications. Sun *et al.* split the functional requirements of an integration process in user interface integration, process integration and data integration, while they classify the key non-functional requirements as security, privacy, billing, and QoS reporting. They then propose SaaS-DL as an extension of Web Service Definition Language (WSDL), and introduce a reference architecture and prototype for a SaaS integration framework. In a case study they found integrating functional requirements possible even using existing SOA techniques, but they note that most SaaS providers do not provide programmatic interfaces to retrieve QoS and billing information which would be necessary to satisfy non-functional integration requirements.

Mikkilineni and Sarathy [39] compared the evolution of cloud computing with Intelligent Network infrastructure in telecommunications and proposed a Virtual Resource Mediation Layer (VRML) to support interoperability between public and private clouds. VRML is an abstraction layer that sits on top of the IaaS layer and allows applications to access CPU, memory, bandwidth and storage depending on needs. The paper fell short of providing any technical details of how such a layer could be implemented, given the APIs used by different IaaS providers are incompatible and disclose only limited information about the real hardware. As mentioned by Grossman [40], vendors are currently resisting standardisation attempts, which make the implementation of such abstraction layers a difficult task.

While much of the research work around cloud interfaces is taking concrete shape, most research on intercloud communication and resource sharing is still focused on defining the research questions and comes without even initial empirical results. This is perhaps to be expected as cloud computing is a relatively new field of research, despite the fact that both the general distributed computing field and the attempt to deliver IT resources as a utility to the consumer has been a goal of research for many decades. So far, a rich intercloud research agenda has been stated by Bernstein *et al.* [34], and it is likely that the search for interoperability will remain a challenging question in cloud computing for a while.

## 7. BUILDING CLOUDS

In this section we describe work that helps building cloud offerings. This requires management software, hardware provision, simulators to evaluate the design, and evaluating management choices.

Sotomayor *et al.* [41] presents two tools for managing cloud infrastructures: OpenNebula, a virtual infrastructure manager, and Haizea, a resource lease manager. To manage the virtual infrastructure, OpenNebula provides a unified view of virtual resources regardless of the underlying virtualisation platform, manages the full lifecycle of the VMs, and support configurable resource allocation policies including policies for times when the demand exceeds the available resources. Sotomayor *et al.* argue that in private and hybrid clouds resources will be limited, in the sense that situations will occur where the demand cannot be met, and that requests for resources will have to be prioritised, queued, pre-reserved, deployed to external clouds, or even rejected. They propose advance reservations to have resources available to serve higher prioritised requests that are expected to be shortly arriving. This can be solved with resource lease managers such as the proposed Haizea, something like a futures market for cloud computing resources, which pre-empts resource usage and puts in place advance resource reservations, so that highly prioritised demand can be served promptly. Haizea can act as a scheduling backend for OpenNebula, and together they advance other virtual infrastructure managers by giving the functionality to scale out to external clouds, and providing support for scheduling groups of VMs, such that either the entire group of VMs are provided resources or no member of the group. In combination they can provide resources by best-effort, as done by Amazon EC2, by immediate provision, as done by Eucalyptus, and in addition using advance reservations.

Song *et al.* [42] have extended IBM data centre management software to be able to deal with cloud-scale data centres, by using a hierarchical set up of management servers instead of a central one. As even simple tasks such as discovering systems or collecting inventory can overwhelm a single management server when the number of managed components or endpoints increases, they partition the endpoints to balance the management workload. Song *et al.* chose a hierarchical distribution of management components, as a centralised topology will in any possible implementation result in bottlenecks, and because P2P structuring exhibits complexities that are not easy to understand. For resilience, the management components have backup servers which are notified with the changes from the original server. Once this notification no longer arrives, the backup server will replace the original server's task until it comes back to operation. In a case study, Song *et al.* show that this solution scales "almost linearly" to 2048 managed endpoints with 8 managing servers. However, cloud-scale solutions might need to manage a number of virtual machines that is one or two orders of magnitude larger, and in the future will become even larger. It is left for future work to test if the solution will be feasible and scale for such numbers of managed endpoints.

Vishwanath *et al.* [43] describe the provision of shipping containers that contain building blocks for data centres. The containers described are not serviced over their lifecycle, but allow for graceful failure of components until performance degrades below a certain threshold and the entire container gets replaced. To achieve this, Vishwanath *et al.* start with over-provisioning the demand at the start, or by putting cold nodes into the container which are not powered on once there is demand due to failure in some of the other components. This work aims at supporting the design of shipping containers with respect to costs, performance, and reliability. For reliability, Markov chains are used to calculate the expected mean time to failure over the lifecycle. For performance and cost, these Markov chains are extended into Markov reward models. These happen under the assumption of exponential failure times, and need to be evaluated against real data. The shipping containers could be used for selling private clouds in a box.

Sriram [44] discusses some of the issues with scaling the size of data centres used to provide cloud computing services. He presents the development and initial results of a simulation tool for predicting the performance of cloud computing data centres which incorporates normal failures, failures that occur frequently due to the sheer number of components and the expected average lifecycle of each component and that are treated as the normal case rather than as an exception. Sriram shows that for small data centres and small failure rates the middleware protocol does not play a role, but for large data centres distributed middleware protocols scale better. CloudSim, another modelling and simulation toolkit has been proposed by Buyya *et al.* [45]. CloudSim simulates the performance of consumer applications executed in the cloud. The topology contains a resource broker and the data centres where the application is executed. The simulator can then estimate the performance overhead of the cloud solution. CloudSim is built on top of a grid computing simulator (GridSim) and looks at the scheduling of the execution application, and the impact of virtualisation on the application's performance.

AbdelSalam *et al.* [46] seek to optimise change management strategies, which are necessary for updates and maintenance, for low energy consumption of a could data centre. However, this work simply derives the actual load from the Service Level Agreements (SLA) negotiated with current customers. AbdelSalam then show that the number of servers currently required is proportional to the load, and identifies the number of idle servers as those available after all SLAs are fulfilled on a minimum set of servers. These are suggested as candidates for pending change management requests. One of the key aspects of cloud computing is elasticity, however, which will make it difficult to estimate the load from the SLAs in place. It is a challenge to develop such placement algorithms that the existing load can always be shrunk to a subset of the available servers while still fulfilling all SLAs, and cost factors will seek to minimise idle servers. Further work is necessary that takes these requirements into account and develops guidelines for both saving energy consumption and enabling seamless change management in cloud data centres.

In summary, several projects research into the way future clouds can be built. Given the methodology we chose earlier, the papers discussed in this section differ too much to conclude with a single research direction in which academia is heading when looking into building future clouds. In fact, it seems there are many more research directions we will be facing when it comes to building new cloud facilities. All papers in this section for example, looked only at IaaS level clouds. To date, no paper could be found that describes technologies for building clouds at another level.

## 8. NEW USE CASES IN CLOUD COMPUTING

In this paper we have so far presented work that seeks to advance the technology of cloud computing. We end this by presenting new technologies and use cases that become possible through the use of cloud computing. Chun and Maniatis [47] describe one such use-case, where cloud computing enables a technology which otherwise would not be possible: to overcome hardware limitations and enable more powerful applications on smartphones, they use external resources. This is done by partially off-loading execution from the smartphone and using cloud resources. But, Chun and Maniatis also include laptops or desktops near the phone in their "cloud" because of the network latency for phones. Depending on the use case, their model offloads entire computations or parts thereof, and only has the remainder executed locally.

Another use-case that becomes feasible and affordable through the use of cloud computing is large-scale non-functional requirements testing, as described by Ganon and Zilbershtein [48]. They tested Network Management Systems for systems where much of the functionality is in the endpoints, such as in voice over IP software. They discuss the advantages and disadvantages of cloud-based testing over testing against real elements or a simulator, and describe how a cloud based test setup can be created using agents that are deployed into the cloud and with the use of cloud elasticity. Further, implications of using the cloud for this setup are evaluated, such as security, safety of intellectual property or software export restrictions, and solutions to tasks such as creating setups that emulate problems including noisy or delayed network connections are presented. Ganon and Zilbershtein reach the conclusion that there are significant benefits of using cloud-based testing, although it cannot completely replace traditional testing against real managed endpoints. They round up their insightful paper with a use case of a test scenario that was carried out on Amazon's cloud and resulted in improvements to the software that could not have been highlighted with other feasible forms of testing, and with disclosing the costs occurred to carry out the cloud based test. Matthew and Spraetz [49] also looked at testing in cloud computing. They explained an effort to automate testing for SaaS providers along the example of Salesforces' Apex test framework. Because consumers can use the Force.com PaaS offering to customise their business solutions into the

CRM system using an entire Java-like programming language, it becomes unfeasible to test all possible states of the CRM beforehand. Instead, a test framework is provided, that allows users to specify regression tests, which can be carried out before every update to the SaaS offering. This is crucial because in a SaaS world there is no choice of version. Once an update is rolled out it is effective for all users.

In a cloud that offers IaaS, the number of VMs and thus instances of operating systems that need to be managed increases significantly. To avoid having to deploy software and updates into each virtual machine, and to avoid lengthy installation processes, entire so called "virtual appliances" will be managed. This means, in cloud computing the operating system will no longer be viewed separated from the applications deployed, but rather both will be deployed and maintained jointly. For service providers this means, they now have the ability to offer a virtual appliance, as functional disc image, instead of having to create lengthy installation procedures to guarantee compatibility with other applications in the VM. Wilson [50] describes *Coronary*, a software configuration management tool for virtual appliances. Coronary takes the idea of incremental updates from configuration management software such as CVS or subversion, and uses this technology to manage virtual appliances over their lifecycle. Wilson discusses the new requirements of version control when used for virtual appliances, and how Coronary handles them.

## 9. CONCLUSION

This paper has presented the work published by the academic community advancing the technology of cloud computing. Much of the work has focussed on creating standards and allowing interoperability, and describes ways of designing and building clouds. We were surprised so far not to see significant contributions to the usage and scaling properties of Hadoop/MapReduce, which is a new programming paradigm in the cloud. Similarly, there was no work published yet on effective usage of PaaS offerings such as Google Apps.

Various definitions of cloud computing were discussed and the NIST working definition by Mell and Grance [11] was found to be the most useful as it described cloud computing using a number of characteristics, service models and deployment models. The socio-technical aspects of cloud computing that were reviewed included the costs of using and building clouds, the security, legal and privacy implications that cloud computing raises as well as the effects of cloud computing on the work of IT departments. The technological aspects that were reviewed included standards, cloud interoperability, lessons from related technologies, building clouds, and use-cases that presented new technological possibilities enabled by the cloud.

A number of authors have discussed the new research challenges that are raised by cloud computing. Bernstein *et al.* [34] listed a research agenda and open questions to achieve interoperability, and Birman *et al.* [29] described a research agenda that seeks to facilitate industry in building successful clouds. Vouk [21] described the problems of managing virtual machine (VM) images. It would be difficult to manually update a large number of VM images and verify their integrity by checking their contents. Mei *et al.* [51] compared the input-output, storage and processing features of cloud computing with pervasive computing and service computing to highlight new research challenges. Cloud computing could benefit from the functionality modelling issues studied in service computing, and the context-sensitivity issues studied in pervasive computing [51]. However, it is difficult to talk about cloud computing without having a particular abstraction layer in mind. The comparisons done by Mei *et al.* are reasonable at an IaaS layer, but they are not very meaningful at the SaaS layer where storage and processing features might not be visible at all. Youseff *et al.* [16] briefly discussed the research challenges in IaaS clouds mentioning that system monitoring information could be used for application optimization in clouds. However, making such information available to users in a useful manner is a challenge [16]. Armbrust *et al.* [18] looked at other research challenges in cloud computing. They highlighted ten obstacles in cloud computing that included technical challenges relating to the adoption of cloud computing, such as availability of service and data lock-in. The lack of scalable storage, performance unpredictability and data transfer bottlenecks are also obstacles that could limit the growth of cloud computing. These obstacles present a number of new research opportunities in cloud computing and Armbrust *et al.* provided some ideas of how these obstacles could be tackled.

To conclude, this paper discussed the research academia has pursued to advance the technological aspects of cloud computing, and highlighted the resulting directions of research facing the academic community. In this way the various projects were set in context, and the research agenda followed by and facing academia was presented. The review showed that there are several ways in which the cloud research community can learn from related communities, and has shown there is interest in academia for describing these similarities. Further, there have been attempts at building unified APIs to access clouds which seem to be more politically than technically challenging. Then, the perhaps clearest research agenda was presented towards interoperability in the cloud and the challenges that need to be overcome. Finally, both for building clouds and presenting use cases in the cloud, the research efforts were shown to be very diverse, making it hard to suggest in which way academia will be moving. This paper reviewed the technical aspects of research in cloud computing. Together with [7], which discussed the work on implications of cloud computing on enterprises and users, this forms a complete survey of all research published on Cloud Computing, providing a solid basis for the $1^{st}$ ACM Symposium on Cloud Computing.

## 10. ACKNOWLEDGMENTS

We thank the Scottish Informatics and Computer Science Alliance (SICSA) and Hewlett-Packard's Automated Infrastructure and Cloud Computing Lab for funding the authors. We also thank the EPSRC for funding the UK's Large-Scale Complex IT Systems (LSCITS) initiative, which enabled our collaboration. We are grateful for the guidance and supervision by Prof. Ian Sommerville and Prof. Dave Cliff, and the entire LSCITS community for their insightful discussions. We also thank Tim Storer for critical reading, and for challenging the results presented here with commonly known solutions in the field of distributed computing.

# REFERENCES

[1] Sun Microsystems, Introduction to Cloud Computing Architecture, 2009

[2] FELLOWS, W. 2008. Partly Cloudy, Blue-Sky Thinking About Cloud Computing. 451 Group.

[3] VARIA, J. 2009. *Cloud Architectures*. Amazon Web Services.

[4] CHAPPELL, D. 2009. *Introducing the Azure Services Platform*. David Chappell & Associates.

[5] RAYPORT, J. F. and HEYWARD, A. 2009. *Envisioning the Cloud: The Next Computing Paradigm*. Marketspace.

[6] PASTAKI RAD, M., SAJEDI BADASHIAN, A., MEYDANIPOUR, G., ASHURZAD DELCHEH, M., ALIPOUR, M. and AFZALI, H. 2009. A Survey of Cloud Platforms and Their Future.

[7] KHAJEH-HOSSEINI, A., SOMMERVILLE, I. and SRIRAM, I. "Research Challenges for Enterprise Cloud Computing," (unpublished).( Submitted to 1st ACM Symposium on Cloud Computing, Indianapolis, Indiana, USA, June 2010, under paper id 54)

[8] IBM, *Staying aloft in tough times*, 2009

[9] PLUMMER, D.C., BITTMAN, T.J., AUSTIN, T., CEARLEY, D.W., and SMITH D.M., *Cloud Computing: Defining and Describing an Emerging Phenomenon*, 2008

[10] STATEN, J., *Is Cloud Computing Ready For The Enterprise?*, 2008.

[11] MELL, P. and GRANCE, T. 2009. Draft NIST Working Definition of Cloud Computing.

[12] ERDOGMUS, H. 2009. Cloud Computing: Does Nirvana Hide behind the Nebula? *Software, IEEE* 26, 2, 4-6.

[13] LEMOS, R. 2009. Inside One Firm's Private Cloud Journey. Retrieved December 1, 2009, from http://www.cio.com/article/506114/Inside_One_Firm_s_Private_Cloud_Journey

[14] Open Cirrus[TM]: the HP/Intel/Yahoo! Open Cloud Computing Research Testbed. Retrieved December 1, 2009, from https://opencirrus.org/

[15] VAQUERO, L., MERINO, L., CACERES, J. and LINDNER, M. 2009. A break in the clouds: towards a cloud definition. *SIGCOMM Comput. Commun. Rev.* 39, 1, 50-55.

[16] YOUSEFF, L., BUTRICO, M. and DA SILVA, D. 2008. Toward a Unified Ontology of Cloud Computing. In *Grid Computing Environments Workshop, 2008. GCE '08*, 1-10.

[17] J. Appavoo, V. Uhlig, and A. Waterland, "Project Kittyhawk: Building a Global-Scale Computer," vol. 42, 2008, pp. pp. 77-84.

[18] M. Armbrust, A. Fox, R. Griffith, A. Joseph, R. Katz, A. Konwinski, G. Lee, D. Patterson, A. Rabkin, I. Stoica, and M. Zaharia, *Above the Clouds: A Berkeley View of Cloud Computing*, 2009.

[19] WEINMAN, J. 2008. 10 Reasons Why Telcos Will Dominate Enterprise Cloud Computing

[20] VOAS, J. and ZHANG, J. 2009. Cloud Computing: New Wine or Just a New Bottle? *IT Professional* 11, 2, 15-17.

[21] VOUK, M. A. 2008. Cloud computing — Issues, research and implementations. In *Information Technology Interfaces, 2008. ITI 2008. 30th International Conference on*, 31-40.

[22] FOSTER, I., ZHAO, Y., RAICU, I. and LU, S. 2008. Cloud Computing and Grid Computing 360-Degree Compared. In *Grid Computing Environments Workshop (GCE '08)*, Austin, Texas, USA, November 2008, 1-10.

[23] Corbató, F. J., Saltzer, J. H., and Clingen, C. T. 1972. Multics: the first seven years. In *Proceedings of the May 16-18, 1972, Spring Joint Computer Conference*, Atlantic City, New Jersey, May 1972, 571-583.

[24] BUYYA, R., YEO, C. and VENUGOPAL, S. 2008. Market-Oriented Cloud Computing: Vision, Hype, and Reality for Delivering IT Services as Computing Utilities. In *High Performance Computing and Communications, 2008. HPCC '08. 10th IEEE International Conference on*, 5-13.

[25] CHANG, M., HE, J., and E. Leon, "Service-Orientation in the Computing Infrastructure," 2006, pp. 27-33.

[26] SEDAYAO, J. 2008. Implementing and operating an internet scale distributed application using service oriented architecture principles and cloud computing infrastructure. In *iiWAS '08: Proceedings of the 10th International Conference on Information Integration and Web-based Applications & Services*, 417-421.

[27] ZHANG, L. and ZHOU, Q. 2009. CCOA: Cloud Computing Open Architecture. In *Web Services, 2009. ICWS 2009. IEEE International Conference on*, 607-616.

[28] NAPPER, J. and BIENTINESI, P. 2009. Can cloud computing reach the top500? In UCHPC-MAW '09: Proceedings of the combined workshops on UnConventional high performance computing workshop plus memory access workshop, 17-20.

[29] Birman, K., Chockler, G., and van Renesse, R. 2009. Toward a cloud computing research agenda. *SIGACT News*, 40, 2, 68-80.

[30] KEAHEY, K., TSUGAWA, M., MATSUNAGA, A. and FORTES, J. 2009. Sky Computing. *Internet Computing, IEEE* 13, 5, 43-51.

[31] NURMI, D., WOLSKI, R., GRZEGORCZYK, C., OBERTELLI, G., SOMAN, S., YOUSEFF, L. and ZAGORODNOV, D. 2008. The Eucalyptus Open-source Cloud-computing System. *Proceedings of Cloud Computing and Its Applications*.

[32] DODDA, R., SMITH, C., and MOORSEL, A. An Architecture for Cross-Cloud System Management, 2009, pp. 556-567.

[33] HARMER, T., WRIGHT, P., CUNNINGHAM, C. and PERROTT, R. 2009. Provider-Independent Use of the Cloud.

[34] BERNSTEIN, D., LUDVIGSON, E., SANKAR, K., DIAMOND, S. and MORROW, M. 2009. Blueprint for the Intercloud - Protocols and Formats for Cloud Computing Interoperability. In *Internet and Web Applications and*

[35] OHLMAN, B., ERIKSSON, A. and REMBARZ, R. 2009. What Networking of Information Can Do for Cloud Computing. In Enabling Technologies: Infrastructures for Collaborative Enterprises, 2009. WETICE '09. 18th IEEE International Workshops on, 78-83.

Services, 2009. ICIW '09. Fourth International Conference on, 328-336.

[36] MATTHEWS, J., GARFINKEL, T., HOFF, C. and WHEELER, J. 2009. Virtual machine contracts for datacenter and cloud computing environments. In *ACDC '09: Proceedings of the 1st workshop on Automated control for datacenters and clouds*, 25-30.

[37] LIM, H., BABU, S., CHASE, J. and PAREKH, S. 2009. Automated control in cloud computing: challenges and opportunities. In *ACDC '09: Proceedings of the 1st workshop on Automated control for datacenters and clouds*, 13-18.

[38] SUN, W., ZHANG, K., CHEN, S.-K., ZHANG, X. and LIANG, H. 2007. Software as a Service: An Integration Perspective.

[39] MIKKILINENI, R. and SARATHY, V. 2009. Cloud Computing and the Lessons from the Past. In Enabling Technologies: Infrastructures for Collaborative Enterprises, 2009. WETICE '09. 18th IEEE International Workshops on, 57-62.

[40] GROSSMAN, R. L. 2009. The Case for Cloud Computing. *IT Professional* 11, 2, 23-27.

[41] SOTOMAYOR, B., RUBÃ©, A., LLORENTE, I. and FOSTER, I. 2009. Virtual Infrastructure Management in Private and Hybrid Clouds. *Internet Computing, IEEE* 13, 5, 14-22.

[42] SONG, S., RYU, K. and DA SILVA, D. 2009. Blue Eyes: Scalable and reliable system management for cloud computing. In *Parallel & Distributed Processing, 2009. IPDPS 2009. IEEE International Symposium on*, 1-8.

[43] VISHWANATH, K., GREENBERG, A. and REED, D. 2009. Modular Data Centers: How to Design Them?

[44] SRIRAM, I. 2009. A Simulation Tool Exploring Cloud-Scale Data Centres. In *1st International Conference on Cloud Computing (CloudCom 2009)*, pp. 381-392.

[45] BUYYA, R., RANJAN, R. and CALHEIROS, R. N. 2009. Modeling and simulation of scalable Cloud computing environments and the CloudSim toolkit: Challenges and opportunities. In *High Performance Computing & Simulation, 2009. HPCS '09. International Conference on*, 1-11.

[46] H. AbdelSalam, K. Maly, R. Mukkamala, M. Zubair, and D. Kaminsky, "Towards Energy Efficient Change Management in a Cloud Computing Environment," 2009, pp. 161-166.

[47] CHUN, B.-G. and MANIATIS, P. 2009. Augmented Smart Phone Applications Through Clone Cloud Execution. In *Proceedings of the 12th Workshop on Hot Topics in Operating Systems (HotOS XII)*.

[48] GANON, Z. and ZILBERSHTEIN, I. E. 2009. Cloud-based Performance Testing of Network Management Systems. In Computer Aided Modeling and Design of Communication Links and Networks, 2009. CAMAD '09. IEEE 14th International Workshop on, 1-6.

[49] MATHEW, R. and SPRAETZ, R. 2009. Test Automation on a SaaS Platform. In Software Testing Verification and Validation, 2009. ICST '09. International Conference on, 317-325.

[50] WILSON, M. 2009. Constructing and Managing Appliances for Cloud Deployments from Repositories of Reusable Components.

[51] MEI, L., CHAN, W. K. and TSE, T. H. 2008. A Tale of Clouds: Paradigm Comparisons and Some Thoughts on Research Issues. In *Asia-Pacific Services Computing Conference, 2008. APSCC '08. IEEE*, 464-469.